\RequirePackage{pdf14}
\documentclass[reprint,
	amsmath,amssymb,
	aps,
	pra,
	floatfix,
]{revtex4-2}

\usepackage{graphicx}\usepackage{dcolumn}\usepackage{bm}\usepackage{hyperref}\usepackage{svg}
	\usepackage{braket}
	\usepackage{mathtools}
	\usepackage{color}
	\usepackage{orcidlink}
	\usepackage{dsfont}
	\usepackage{overpic}
	\usepackage{siunitx}
    \usepackage{cancel}

\usepackage{soul}

\hypersetup{
 colorlinks = true,
 linkcolor = {blue},
 citecolor = {blue},
 urlcolor = {blue},
}

\usepackage[whole]{bxcjkjatype}
\usepackage[normalem]{ulem}

\begin{document}

\title{Active interference suppression in frequency-division-multiplexed quantum gates \\via off-resonant microwave tones
}

\author{Haruki Mitarai\orcidlink{0000-0002-8374-3648}}
 \email{hmitarai@mosk.tytlabs.co.jp}\author{Yukihiro Tadokoro\orcidlink{0000-0002-1844-3040}}\email{y.tadokoro@ieee.org}
\author{Hiroya Tanaka\orcidlink{0000-0002-5113-0927}}\email{tanak@mosk.tytlabs.co.jp}

\affiliation{Toyota Central R\&D Labs., Inc., 41-1, Yokomichi, Nagakute, Aichi 480-1192, Japan}

\date{\today}

\begin{abstract}
	The increasing number of control lines connecting quantum processors to external electronics constitutes a major bottleneck in the realization of large-scale quantum computers.
    Frequency-division multiplexing is expected to enable control of multiple qubits through a single microwave cable; 
    however, interference from off-resonant microwave tones hinders precise qubit control.
	Here, we propose an active interference suppression method for frequency-division-multiplexed simultaneous gates on microwave-controlled qubits. 
	We demonstrate that the deliberate incorporation of
	off-resonant microwave tones improves single-qubit gate fidelity. 
In particular, the gate infidelity scales inversely with the square of the number of microwave tones when off-resonant orthogonal or quasi-orthogonal tones are incorporated.
Furthermore, we show that fast oscillations, neglected under the rotating wave approximation, degrade the gate fidelity, and that this degradation can be mitigated through optimized frequency allocation.
	The proposed approach is simple and effective for improving the performance of frequency-division-multiplexed quantum gates.
\end{abstract}

\maketitle

\section{Introduction}
	In recent years, significant efforts have been directed toward developing fault-tolerant quantum computers~\cite{nakamura1999coherent, blais2004cavity, Charge-insensitive, demonstration, IBM_qubit, PhysRevLett.129.030501, bluvstein2024logical, PRXQuantum.5.030347, optical_QC}, which have potential applications in prime factorization~\cite{Shor1}, linear system solvers~\cite{Harrow}, and computer-aided engineering~\cite{sato_CAE}. 
	Among the various available platforms, superconducting circuits are highly promising; notably, quantum supremacy and error correction based on  surface codes have been experimentally demonstrated~\cite{google, google2025quantum, Surface_codes_hefei}. 
	However, achieving quantum advantage in practical applications requires millions of qubits~\cite{Surface_codes, gameofsurfacecodes}. 
	Building a quantum computer at such scales demands not only improved qubit performance but also substantial advances in the associated control electronics. 

	Currently, the control architectures for solid-state quantum processors commonly generate microwave signals at room temperature~\cite{GaAs_qubit, refrigerator, VANDIJK201990,  Si_qubit_cmos, takeuchi2024microwave}. 
	These signals are routed to individual qubits in a dilution refrigerator using one-to-one coaxial cables. 
	However, this wiring model is not scalable, as the number of required control wires increases with the number of qubits.
	Increasing the number of control wires adds hardware complexity and costs. Moreover, additional heat loads are introduced through physical interconnections, thereby threatening the cooling capacity of the dilution refrigerator~\cite{refrigerator}. 
	
	Hence, reducing the number of control wires is critical for scaling quantum computing systems~\cite{Baseband, PhysRevResearch.5.013145, IOP_cryo_2023, ohira2024optimizing, Roberto_May2025}. 
	Frequency-division multiplexing (FDM) enables a single microwave cable to control multiple qubits simultaneously~\cite{Cryo-CMOS_FDMA, PhysRevApplied.18.064046, JJ_SDM_FDM}, thereby reducing the number of required control wires. 
	However, a microwave tone intended for a qubit is off-resonant with other simultaneously driven qubits and can induce unwanted crosstalk, which hinders precise parallel gate operations~\cite{Impact_of, ohira2024optimizing}.
	Suppressing these crosstalk effects is therefore essential for achieving high-fidelity gates in FDM schemes. 
	Several studies have combined FDM with cryogenic qubit controllers~\cite{Cryo-CMOS_FDMA, PhysRevApplied.18.064046, takeuchi2024microwave}. 
	For example, a frequency-division-multiplexed qubit controller based on adiabatic quantum-flux-parametron logic has been previously demonstrated~\cite{takeuchi2024microwave}. 
	However, developing cryogenic qubit controllers remains technically challenging. 
	Recent studies have explored pulse spectral engineering, which can circumvent these challenges~\cite{selective_excitation, OFDM_gate}. 
	For instance, selective gate operations enabled by pulse shaping have been demonstrated~\cite{selective_excitation}. 

	In this study, we propose an active interference suppression (AIS) method for simultaneous gate operations in FDM schemes. 
	The proposed method is based on the recent observation that the incorporation of appropriately selected off-resonant tones can suppress interference~\cite{OFDM_gate}. 
	We show that the deliberate incorporation of off-resonant orthogonal or quasi-orthogonal microwave tones improves the single-qubit gate fidelity. 
	Specifically, 
	for qubits which are arranged at regular intervals in the frequency domain, the gate infidelity scales inversely with the square of the number of applied microwave tones.
	We further demonstrate that fast-oscillating terms neglected under the rotating wave approximation (RWA) affect the gate fidelity in the AIS method.
	By shifting the central frequency of the off-resonant microwave tones, we compensate for this effect and achieve higher fidelity.

	AIS offers an alternative perspective to the prevailing notion that off-resonant microwave tones are detrimental to qubit control. 
	Our results are consistent with those of recent studies that highlight the constructive role of carefully engineered off-resonant components~\cite{OFDM_gate, ALC}. 
	Moreover, the applicability of the proposed framework is not limited to FDM-based simultaneous gate operations.
	The method can be readily extended to other tasks in qubit manipulation,
	such as selective excitation~\cite{selective_excitation} and microwave crosstalk mitigation~\cite{microwave_crosstalk_APL}.

\section{Model} \label{sec:model}
	\subsection{System Hamiltonian} \label{sec:model:hamiltonian}
		\begin{figure}[t]
			\centering
			\includegraphics[width = 3.4in]{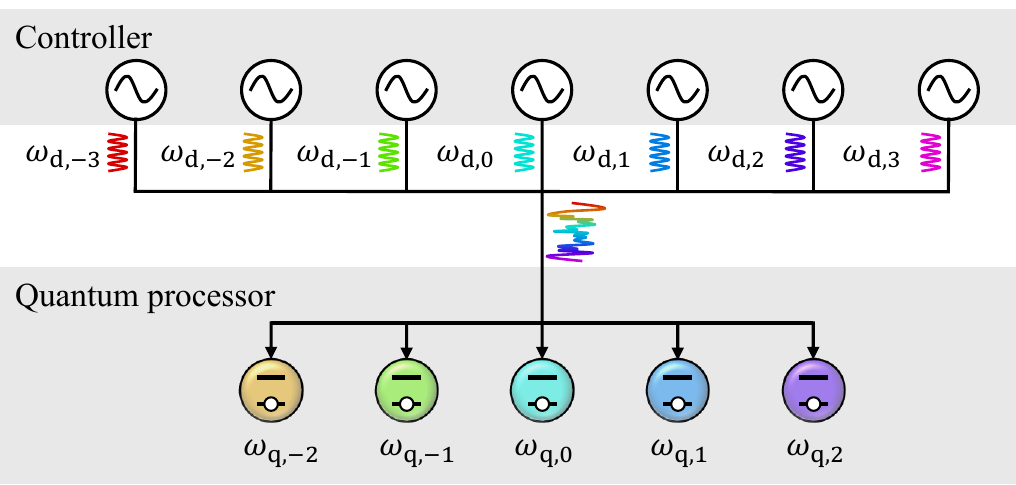}
			\caption{Conceptual illustration of a model comprising $N_{\mathrm{q}}$ independent qubits with transition frequencies $\omega_{\mathrm{q}, k_{\mathrm{q}}}$, driven by microwaves via a shared control line. 
			The controller produces $N_{\mathrm{d}}$ microwave tones at distinct frequencies $\omega_{\mathrm{d}, k_{\mathrm{d}}}$, which are then combined and routed to a quantum processor via the shared line. 
			The combined signal is applied uniformly to all qubits. 
			In this figure, $N_{\mathrm{q}}=5$ and $N_{\mathrm{d}} = 7$. 
			}
			\label{fig:model}
		\end{figure}
	
\begin{figure}[t]
			\centering
			\includegraphics[width = 3.4in]{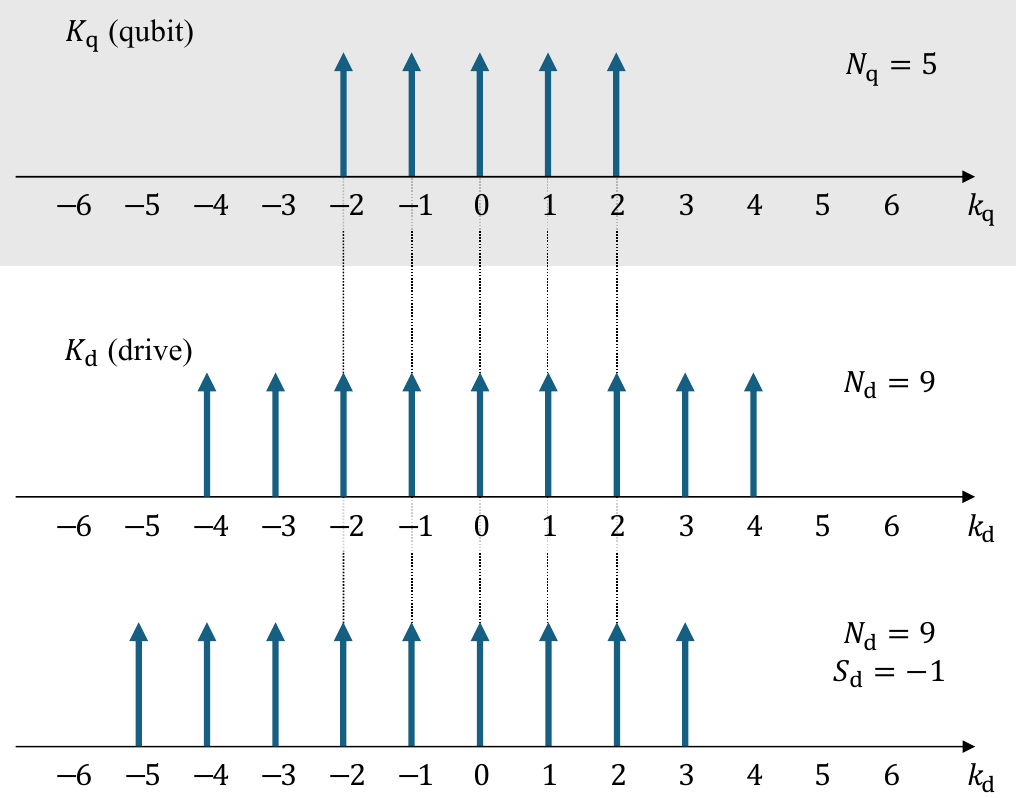}
			\begin{picture}(700, 0)
				\put(2, 193){(a)}
				\put(2, 116){(b)}
				\put(2, 55){(c)}
			\end{picture}
			\caption{Illustration of the set $K_{\mathrm{q}}$ and $K_{\mathrm{d}}$.
			(a) Set of qubit indices $K_{\mathrm{q}}$ for $N_{\mathrm{q}} = 5$.
			(b), (c) Sets of indices of drive microwave tones $K_{\mathrm{d}}$ for $N_{\mathrm{d}} = 9$. 
The set $K_{\mathrm{d}}$ shown in (b) corresponds to symmetric allocation, i.e., $S_{\mathrm{d}} = 0$.   
			The set shown in (c) corresponds to weakly asymmetric allocation with $S_{\mathrm{d}} = -1$. 
}
			\label{fig:K}
		\end{figure}

		Figure~\ref{fig:model} illustrates a system of $N_{\mathrm{q}}$ independent qubits driven by a frequency-division-multiplexed microwave signal.
		The controller generates $N_{\mathrm{d}}$ microwave tones, which are combined and transmitted to a quantum processor through a shared control line. 
		The synthesized signal is applied uniformly to all the qubits.
		The Hamiltonian of this system is given by 
\begin{align}
			\label{eq:H_all}
			H_{\text{total}}\left(t\right)
			\equiv \sum_{k_{\mathrm{q}} \in K_{\mathrm{q}}} H_{k_\mathrm{q}}, 
		\end{align}
        with single-qubit Hamiltonian
		\begin{align}
			\label{eq:H}
			H_{k_\mathrm{q}}\left(t\right) \equiv - & \frac{\omega_{\mathrm{q}, k_{\mathrm{q}}}}{2} \sigma_{\mathrm{z}, k_{\mathrm{q}}} \nonumber \\
			& + \sum_{k_{\mathrm{d}} \in K_{\mathrm{d}}}  \alpha_{k_{\mathrm{d}}} s\left(t\right) \sin\left(\omega_{\mathrm{d}, k_{\mathrm{d}}} t\right) \sigma_{\mathrm{y}, k_{\mathrm{q}}}, 
		\end{align}
		where $\omega_{\mathrm{d}, k_{\mathrm{d}}}$ and $\alpha_{k_{\mathrm{d}}}$ denote the frequency and amplitude of the $k_{\mathrm{d}}$th microwave tone, respectively. 
        The index $k_{\mathrm{d}}$, which labels the drive microwave tones, is a member of the set $K_{\mathrm{d}}$ defined by
        \begin{align*}
            K_{\mathrm{d}} \equiv &\left\{-\lfloor N_{\mathrm{d}} / 2\rfloor, -\lfloor N_{\mathrm{d}} / 2\rfloor + 1, \right. \\
            &\left. \dots, \lfloor \left(N_{\mathrm{d}}-1\right) / 2\rfloor - 1, \lfloor \left(N_{\mathrm{d}}-1\right)/ 2\rfloor\right\} + S_{\mathrm{d}},
        \end{align*}
        where, for a set $A$ and an integer $b$, $A + b$ denotes the translated set $\left\{a + b \mid a \in A\right\}$. 
        The shift parameter $S_{\mathrm{d}}$ is introduced to allow asymmetric allocation. 
        The transition frequency of the $k_{\mathrm{q}}$th qubit is $\omega_{\mathrm{q}, k_{\mathrm{q}}}$, 
        where $k_{\mathrm{q}} \in K_{\mathrm{q}}$ is the qubit index. 
        Here, the set $K_{\mathrm{q}}$ is defined by
        \begin{align*}
            K_{\mathrm{q}} \equiv &
            \left\{-\left(N_{\mathrm{q}} - 1\right)/2, -\left(N_{\mathrm{q}} - 1\right)/2 + 1, \right. \\ 
            &\quad \left. \dots, \left(N_{\mathrm{q}} - 1\right)/2-1, \left(N_{\mathrm{q}} - 1\right)/2\right\}
        \end{align*}
for odd $N_{\mathrm{q}}$. 
        In what follows, we consider only odd $N_{\mathrm{q}}$ for convenience.  
        Figure~\ref{fig:K}(a) shows an example of $K_{\mathrm{q}}$ for $N_{\mathrm{q}} = 5$.
The operators $\sigma_{\mathrm{x}, k_{\mathrm{q}}}$, $\sigma_{\mathrm{y}, k_{\mathrm{q}}}$, and $\sigma_{\mathrm{z}, k_{\mathrm{q}}}$ are the Pauli operators associated with the $k_{\mathrm{q}}$th qubit. 
		The function $s\left(t\right)$ represents the pulse envelope. 
		Since arbitrary single-qubit gates can be implemented using $X_{\pi/2}$ and virtual $Z$ gates~\cite{virtual_Z}, we consider only the driving terms involving $\sigma_{\mathrm{y}, k_{\mathrm{q}}}$.

        We now focus on the $k_{\mathrm{q}}$th qubit and analyze its dynamics using Eq.~\eqref{eq:H}. 
        We allocate qubit frequencies $\omega_{\mathrm{q}, k_{\mathrm{q}}}$ and drive frequencies $\omega_{\mathrm{d}, k_{\mathrm{d}}}$ at regular intervals $\Delta$, where $\Delta \ll \omega_{\mathrm{q}, k_{\mathrm{q}}}, \omega_{\mathrm{d}, k_{\mathrm{d}}}$, 
		and select these frequencies such that $\omega_{\mathrm{q}, k_{\mathrm{q}}} = \omega_{\mathrm{d}, k_{\mathrm{d}}}$ for $k_{\mathrm{q}} = k_{\mathrm{d}}$. 
		This yields $\omega_{\mathrm{d}, k_{\mathrm{d}}} = \omega_{\mathrm{q}, k_{\mathrm{q}}} + \left(k_{\mathrm{d}} - k_{\mathrm{q}}\right)\Delta$.
In addition, we assume that $s \left(t\right)$ is constant when the control is active, and set $\alpha_{k_{\mathrm{d}}} = \alpha$ for all $k_{\mathrm{d}}$. 
		Under these assumptions, 
		the system Hamiltonian in the frame rotating at the qubit frequency is given by 
		\begin{align}
			\label{eq:HI}
			&H_{\mathrm{I}, k_{\mathrm{q}}} \left(t\right) = \frac{\alpha}{2} s\left(t\right)
			\nonumber \\ \quad &
				\sum_{k_{\mathrm{d}} \in K_{\mathrm{d}}}  \left[
				\sigma_{\mathrm{x}, k_{\mathrm{q}}} \left\{\cos 
				\left( \left(2\omega_{\mathrm{q}, k_{\mathrm{q}}} + \Delta_{k_{\mathrm{d}} k_{\mathrm{q}}} \right)t\right) - \cos \left(\Delta_{k_{\mathrm{d}} k_{\mathrm{q}}} t \right)\right\} \right.\nonumber\\
				\quad &\left.+ \sigma_{\mathrm{y}, k_{\mathrm{q}}} \left\{\sin \left(\left(2\omega_{\mathrm{q}, k_{\mathrm{q}}} + \Delta_{k_{\mathrm{d}} k_{\mathrm{q}}}\right)t\right) + \sin \left( \Delta_{k_{\mathrm{d}} k_{\mathrm{q}}} t \right)\right\}
			\right] ,
		\end{align}
		where $\Delta_{kj} \equiv (k-j)\Delta$, 
		\begin{align}
			\label{eq:s}
			s\left(t\right) = \left\{ \,
				\begin{aligned}
					&1, \quad 0 \le t \le \tau,\\
					&0, \quad \text{otherwise},
				\end{aligned}
				\right.
		\end{align}
		and $\tau$ denotes the pulse width. 
		Equation~\eqref{eq:HI} can be simplified by applying the RWA:
		\begin{align}
			\label{eq:HI_RWA}
			H_{\mathrm{I}, k_{\mathrm{q}}} \left(t\right)
			 & \sim \frac{\alpha}{2} s \left(t\right) \nonumber \\ & \sum_{k_{\mathrm{d}} \in K_{\mathrm{d}}} \left[
			- \sigma_{\mathrm{x}, k_{\mathrm{q}}} \cos \left( \Delta_{k_{\mathrm{d}} k_{\mathrm{q}}} t \right)
			+ \sigma_{\mathrm{y}, k_{\mathrm{q}}} \sin \left( \Delta_{k_{\mathrm{d}} k_{\mathrm{q}}} t \right) \right]. 
		\end{align}
		Equation~\eqref{eq:HI_RWA} is used in the theoretical analysis presented in Sec.~\ref{sec:method}. 

	\subsection{Orthogonal and quasi-orthogonal pulses} \label{sec:model:pulse}
		\begin{figure}[tb]
		\centering
		\includegraphics[scale=1]{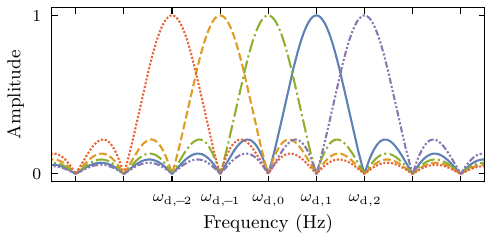}
		\caption{Absolute values of the normalized spectra of $s\left(t\right) \sin\left(\omega_{\mathrm{d},  k_{\mathrm{d}}} t\right)$ for $k_{\mathrm{d}} \in \left\{-2, -1, 0, 1, 2\right\}$ with $\tau = \tau_0$. 
			Each drive frequency $\omega_{\mathrm{d}, k_{\mathrm{d}}}$ is located at a zero crossing of the spectral components of all the other microwave tones.
		}
		\label{fig:spectrum}
	\end{figure}
		For the Hamiltonian given in Eq.~\eqref{eq:HI}, the gate fidelity improves at specific values of the pulse width $\tau$. 
		In this context, we introduce orthogonal and quasi-orthogonal pulses, which are defined in terms of the pulse width $\tau$~\cite{OFDM_gate}. 
		
		The Fourier transform of the pulse $s\left(t\right) \sin\left(\omega_{\mathrm{d}, k_{\mathrm{d}}}t\right)$ in Eq.~\eqref{eq:H} is proportional to a sinc function, which has zero crossings at regular intervals of $2\pi/\tau$, except at its center. 
		For $\tau = \tau_0$, where $\tau_0 \equiv 2\pi/\Delta$, the spacing between adjacent zero crossings coincides with the drive and qubit frequency intervals $\Delta$. 
		Consequently, as shown in Fig.~\ref{fig:spectrum}, the spectral peak of each microwave tone aligns with the nulls of all the other tones, thereby suppressing interference and yielding high gate fidelity~\cite{OFDM_gate}. 
		This mechanism is conceptually analogous to orthogonal frequency-division multiplexing (OFDM)~\cite{OFDM, OFDM2009, 5G}, which is widely used in wireless communication systems.
Ref.~\cite{OFDM_gate} further showed that high gate fidelity is also achievable for $\tau = \tau_0/2$.

		Similar fidelity improvements can be obtained for nonzero integer multiples of these pulse widths. 
		Let $m$ be a positive integer.
		We refer to pulses with widths $\tau = m\tau_0/2$ as ``orthogonal'' when $m$ is even and ``quasi-orthogonal'' when $m$ is odd.

\section{Theory of active interference suppression} \label{sec:method}
	\begin{figure}[t]
			\centering
			\includegraphics[width = 3.4in]{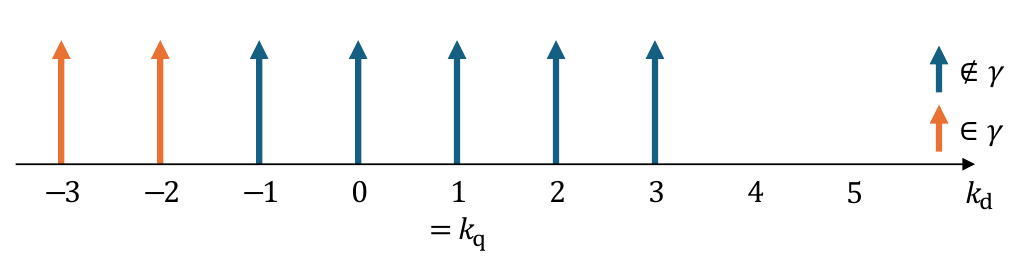}
		\caption{Illustration of $\gamma$, which represents the set of indices of drive microwave tones lacking a corresponding element with respect to $k_{\mathrm{d}} = k_{\mathrm{q}}$. 
		The orange and blue arrows denote the drive microwave tones that correspond and do not correspond to the set $\gamma$, respectively.
		In this figure, $k_{\mathrm{q}} = 1$, $N_\mathrm{d}=7$, and $S_\mathrm{d}=0$, yielding $\gamma = \left\{ k_{\mathrm{q}}-4, k_{\mathrm{q}}-3 \right\} = \left\{ -3, -2 \right\}$. These indices have no counterpart with respect to $k_{\mathrm{q}}$, i.e., there are no arrows at $k_{\mathrm{q}}+3 = 4$ and $k_{\mathrm{q}}+4 = 5$.
		}
		\label{fig:gamma}
	\end{figure}
	
The average gate fidelity $F\left(U_1, U_2\right)$ is defined as follows~\cite{fidelity}:
	\begin{align}
		\label{eq:average_gate_fidelity}
		F\left(U_1, U_2\right) \equiv \frac{\left| \text{Tr}\left(U_{1}^{\dagger}U_2\right)\right|^2 + d}{d\left(d + 1\right)}, 
	\end{align}
	where $U_{1}$ and $U_{2}$ are unitary operators, and $d$ is the dimension of the Hilbert space.
To quantify the impact of interference on gate performance, we use the average gate infidelity $1 - F\left(U_{\text{ideal}}, U\right)$, where 
	$U_{\text{ideal}} \equiv e ^ {-i\frac{\phi}{2} \sigma_{\mathrm{x}, k_{\mathrm{q}}}}$ denotes the desired unitary operator with the rotation angle $\phi$. 
	The operator $U$ denotes the actual evolution operator, which is defined as $\Ket{\psi_{\mathrm{I}}\left(\tau\right)} = U\Ket{\psi_{\mathrm{I}}\left(0\right)}$. 
	
	For an analytical estimate of the infidelity, 
we expand the time-evolution operator using the Magnus expansion~\cite{Magnus}. 
We define the resulting approximate operator $U_{\text{Magnus}}$ as follows:
\begin{align}
		\label{eq:Magnus}
		U_{\text{Magnus}}\left(t\right) \equiv \exp\left[-i\left\{\Omega_1\left(t\right) + \Omega_2\left(t\right)\right\}\right], 
	\end{align}
	where $\Omega_1\left(t\right)$ and $\Omega_2\left(t\right)$ are the first and second terms of the Magnus series, respectively.
	Using Eq.~\eqref{eq:HI_RWA}, $\Omega_1\left(t\right)$ and $\Omega_2\left(t\right)$ take the following forms~\cite{OFDM_gate}: 
	\begin{align}
		\label{eq:Omega_1_1}
		\Omega_1\left(\tau\right) 
		\sim - \left( \sigma_{\mathrm{x}, k_{\mathrm{q}}} \lambda_{\mathrm{x}} +  \sigma_{\mathrm{y}, k_{\mathrm{q}}} \lambda_{\mathrm{y}}\right), 
	\end{align}
	\begin{align}
		\label{eq:Omega_2_1}
		\Omega_2\left(\tau\right) \sim  - \sigma_{\mathrm{z}, k_{\mathrm{q}}} \lambda_{\mathrm{z}},  
	\end{align}
	where $\lambda_{\mathrm{x}}$, $\lambda_{\mathrm{y}}$, and $\lambda_{\mathrm{z}}$ are real coefficients. 
	Using Eqs.~\eqref{eq:Magnus}--\eqref{eq:Omega_2_1}, the infidelity becomes~\cite{OFDM_gate}
	\begin{align}
		\label{eq:F_analytical}
		&1 - F\left(U_{\text{ideal}}, U_{\text{Magnus}}\left(\tau\right)\right) \nonumber \\
		&\; \sim 1- \frac{1}{3} \left( 2 \left| \cos\left|\frac{\phi}{2}\right| \cos\Lambda - \frac{\phi \lambda_{\mathrm{x}}}{\left|\phi\right|\Lambda} \sin\left|\frac{\phi}{2}\right| \sin \Lambda \right|^2 + 1 \right), 
	\end{align}
	where $\Lambda \equiv \sqrt{\lambda_{\mathrm{x}}^2 + \lambda_{\mathrm{y}}^2 + \lambda_{\mathrm{z}}^2}$. 
	Thus, $\lambda_{\mathrm{x}} \sim -\phi/2$ and $\lambda_{\mathrm{y}}, \lambda_{\mathrm{z}} \sim 0$ lead to high-fidelity gates.

	We now introduce the AIS method and theoretically analyze its effects on gate infidelity. 
For orthogonal or quasi-orthogonal pulses, i.e., $\tau = m \tau_0 / 2$, the coefficients $\lambda_{\mathrm{x}}$, $\lambda_{\mathrm{y}}$, and $\lambda_{\mathrm{z}}$ become~\cite{OFDM_gate}
	\begin{align}
		\label{eq:lambda_x_0}
		\lambda_{\mathrm{x}} &= -\frac{\phi}{2}, \\
		\label{eq:lambda_y_0}
		\lambda_{\mathrm{y}} &= - \frac{\phi}{2 \pi m } \sum_{k_{\mathrm{d}} \in \gamma} \frac{\left(-1\right)^{m\left(k_{\mathrm{d}} - k_{\mathrm{q}}\right)} - 1}{k_{\mathrm{d}}-k_{\mathrm{q}}}, \\
		\label{eq:lambda_z_0}
		\lambda_{\mathrm{z}} &= \frac{\phi^2}{4\pi m} \sum_{k_{\mathrm{d}} \in \gamma} \frac{\left(-1\right)^{m\left(k_{\mathrm{d}} - k_{\mathrm{q}}\right)}}{k_{\mathrm{d}}-k_{\mathrm{q}}}. 
	\end{align}
	Here, we set $\alpha = -\phi/\tau$, and $\gamma \subseteq K_{\mathrm{d}}$ denotes the set of indices of the drive microwave tones lacking a corresponding element with respect to $k_{\mathrm{d}} = k_{\mathrm{q}}$ (see Fig.~\ref{fig:gamma}). 
	We clearly have $\lambda_{\mathrm{x}} = - \phi/2$. 
	However, nonzero $\lambda_{\mathrm{y}}$ and $\lambda_{\mathrm{z}}$ degrade the fidelity. 
In particular, gate operations on qubits near the left or right ends [$k_{\mathrm{q}}\sim \pm\left(N_{\mathrm{q}} - 1\right)/2$] show lower fidelity due to the large cardinality~$\left|\gamma\right|$. 

	To compensate for this reduction in fidelity, we introduce additional off-resonant drive tones. 
	In Eqs.~\eqref{eq:lambda_y_0} and \eqref{eq:lambda_z_0}, the coefficients $\left|\lambda_\mathrm{y}\right|$ and $\left|\lambda_\mathrm{z}\right|$ take smaller values when (1) the cardinality $\left|\gamma\right|$ is small or (2) the members of $\gamma$ are far from the qubit index $k_{\mathrm{q}}$. 
	While condition (1) cannot be satisfied simultaneously across all qubits, condition (2) can be fulfilled by adding off-resonant tones symmetrically relative to the central qubit's frequency $\omega_{\mathrm{q},0}$.
Upon selecting $S_{\mathrm{d}} = 0$, which corresponds to $K_{\mathrm{d}}$ shown in Fig.~\ref{fig:K}(b), 
	$\left|\lambda_{\mathrm{y}} \right|$ and $\left|\lambda_{\mathrm{z}} \right|$ for $N_{\mathrm{d}} \gg \left|k_{\mathrm{q}}\right|$ become
	\begin{align}
		\label{eq:lambda_y_0_approx}
		&\left|\lambda_{\mathrm{y}} \right| \sim \frac{\left| \phi \right|}{\pi m N_{\mathrm{d}}} \left| \left| \gamma \right| - G\left(k_{\mathrm{q}}, m, \gamma\right)\right|, \\
		\label{eq:lambda_z_0_approx}
		&\left|\lambda_{\mathrm{z}} \right| \sim \frac{\phi^2}{2\pi m N_{\mathrm{d}}} \left| G\left(k_{\mathrm{q}}, m, \gamma\right) \right|,
	\end{align}  
	where 
	\begin{align}
		G\left(k_{\mathrm{q}}, m, \gamma\right) = \sum_{k_{\mathrm{d}} \in \gamma}\left(-1\right)^{m\left(k_{\mathrm{d}} - k_{\mathrm{q}}\right)}. 
	\end{align} 
	Assuming $\lambda_{\mathrm{x}} = -\phi/2$ and $\left|\lambda_{\mathrm{y}}\right|,\left|\lambda_{\mathrm{z}}\right| \ll \left|\lambda_{\mathrm{x}}\right|$, 
	Eq.~\eqref{eq:F_analytical} reduces to
	\begin{align}
		\label{eq:F_analytical_approx}
		1 - F\left(U_{\text{ideal}}, U_{\text{Magnus}}\left(\tau\right)\right) \sim \frac{\lambda_{\mathrm{y}}^2 + \lambda_{\mathrm{z}}^2}{3\lambda_{\mathrm{x}}^2}. 
	\end{align}
	Thus, substituting Eqs.~\eqref{eq:lambda_x_0},~\eqref{eq:lambda_y_0_approx}, and \eqref{eq:lambda_z_0_approx} into Eq.~\eqref{eq:F_analytical_approx},
	the infidelity can be rewritten as
	\begin{align}
		\label{eq:F_analytical_approx_2}
		1 -  F&\left(U_{\text{ideal}}, U_{\text{Magnus}}\left(\tau\right)\right) \nonumber \\
		& \sim \frac{1}{3 \pi^2 m^2 N_{\mathrm{d}}^2} \left[ 4\left\{ \left| \gamma \right| - G\left(k_{\mathrm{q}}, m, \gamma\right) \right\}^2 \right. \nonumber \\
		&\quad\quad\quad\quad\quad\quad\quad\quad\quad \left. + \phi^2 G^2\left(k_{\mathrm{q}}, m, \gamma\right) \right]. 
	\end{align}
	Here, $G\left(k_{\mathrm{q}}, m, \gamma\right) = \left| \gamma \right|$ for an even $m$, 
	whereas for an odd $m$, it takes the values $0, -1, \text{or}\, 1$.
Additionally, $\left| \gamma \right|$ differs by only one, depending on whether $N_{\mathrm{d}}$ is even or odd.
	Therefore, fixing $m$, $k_{\mathrm{q}}$, and $S_{\mathrm{d}}$, the infidelity is expected to scale as $N_{\mathrm{d}}^{-2}$ when orthogonal or quasi-orthogonal microwave tones are incorporated.

\section{Numerical results} \label{sec:result}
	\begin{figure}[t]
		\centering
		\includegraphics[scale=1]{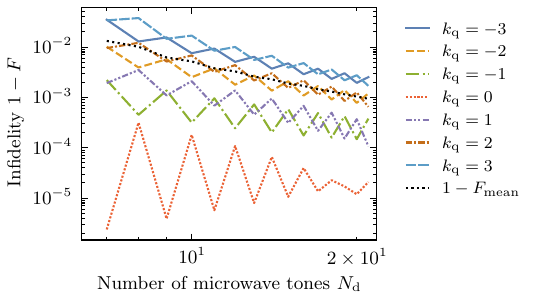}
		\caption{Average gate infidelity $1 - F\left(U_{\text{ideal}}, U\right)$ as a function of the number of drive microwave tones $N_{\mathrm{d}}$ for $\tau = \tau_0$.
		Parameters are set as $N_{\mathrm{q}} = 7$, $S_{\mathrm{d}}=0$, $\omega_{\mathrm{q}, 0} / 2\pi = 5 \,\mathrm{GHz}$, $\Delta/2\pi = 10 \,\mathrm{MHz}$, $\phi = \pi / 2$, and $\alpha = -\phi/\tau$.
		The black dotted line indicates $1 - F_{\text{mean}}\left(U_{\text{ideal}}, U\right)$, which is the mean value of $1 - F\left(U_{\text{ideal}}, U\right)$ over $k_{\mathrm{q}}\in K_{\mathrm{q}} = \left\{-3, -2, -1, 0, 1, 2, 3\right\}$. 
		The operator $U$ is obtained numerically from the Hamiltonian in Eq.~\eqref{eq:H}. 
		}
		\label{fig:result:N}
	\end{figure}
	\begin{figure*}[tb]
		\begin{tabular}{cc}
			\begin{minipage}[t]{4.0in}
				\centering
				\includegraphics[scale=1]{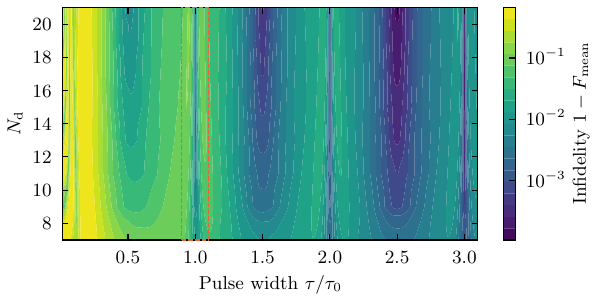}
			\end{minipage}
			\begin{minipage}[t]{3in}
				\centering
				\includegraphics[scale=1]{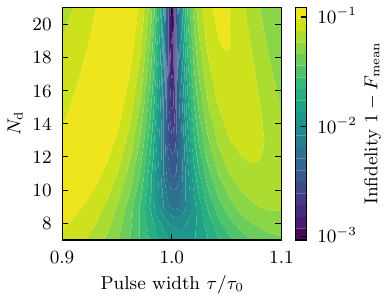}
			\end{minipage}
		\end{tabular}
		\begin{picture}(700, 0)
			\put(4, 140){(a)}
			\put(310, 140){(b)}
		\end{picture}
		\caption{
			(a) Infidelity $1 - F_{\text{mean}}\left(U_{\text{ideal}}, U\right)$ as a function of the number of drive microwave tones $N_\mathrm{d}$ and the normalized pulse width $\tau/\tau_0$ 
			for $0.01 \le \tau/\tau_0 \le 3.1$. 
			(b) Enlarged view of (a) in the range $0.9 \le \tau/\tau_0 \le 1.1$, corresponding to the region outlined by the orange dashed lines.
			Parameters are as follows: $N_{\mathrm{q}} = 7$, $S_{\mathrm{d}}=0$, $\omega_{\mathrm{q}, 0} / 2\pi = 5 \,\mathrm{GHz}$, $\Delta/2\pi = 10 \,\mathrm{MHz}$, $\phi = \pi / 2$, and $\alpha = -\phi/\tau$.
			$F_{\text{mean}}\left(U_{\text{ideal}}, U\right)$ denotes the mean of $F\left(U_{\text{ideal}}, U\right)$ over all qubits $k_{\mathrm{q}} \in K_{\mathrm{q}} = \left\{-3, -2, -1, 0, 1, 2, 3\right\}$. 
			The operator $U$ is derived numerically from the Hamiltonian in Eq.~\eqref{eq:H}. 
			}
		\label{fig:result:tau}
	\end{figure*}
	
In this section, the effectiveness of the AIS method introduced in Sec.~\ref{sec:method} is validated. 
	We also identify the pulse width conditions required by AIS.
	For this purpose, we obtain the exact evolution operator $U$ numerically from the Hamiltonian in Eq.~\eqref{eq:H}.
\subsection{Reduction of gate infidelity with AIS} \label{sec:result:Nd}
		Figure~\ref{fig:result:N} shows the average gate infidelity $1 - F\left(U_{\text{ideal}}, U\right)$ for each qubit as a function of the number of drive microwave tones $N_{\mathrm{d}}$.
		We set $\tau = \tau_0$, $S_{\mathrm{d}}=0$, and $N_{\mathrm{q}} = 7$. 
The black dotted line represents $1 - F_{\text{mean}}\left(U_{\text{ideal}}, U\right)$, which is the mean infidelity over all qubits $k_{\mathrm{q}} \in K_{\mathrm{q}}$.
		Here, $F_{\text{mean}}\left(U_1, U_2\right) \equiv \frac{1}{N_{\mathrm{q}}}\sum_{k_{\mathrm{q}} \in K_{\mathrm{q}}} F\left(U_1, U_2\right)$.
In Fig.~\ref{fig:result:N}, the infidelity decreases with increasing $N_{\mathrm{d}}$. 
Fitting a linear function to the black dotted line in the range $11 \le N_{\mathrm{d}} \le 21$, we obtain a slope of approximately $-2.2$. 
		This result is consistent with the scaling predicted using Eq.~\eqref{eq:F_analytical_approx_2}, indicating that the infidelity scales as $N_{\mathrm{d}}^{-2}$.
		Here, the data for $N_{\mathrm{d}} < 11$ are excluded because Eq.~\eqref{eq:F_analytical_approx_2} assumes $N_{\mathrm{d}} \gg \left|k_{\mathrm{q}}\right|$.
		The jagged features of the infidelity curves reflect the changes in $\left|\gamma\right|$, which differs by one between even and odd values of $N_{\mathrm{d}}$.

		Note that, for $k_{\mathrm{q}} = 0$, the theoretical infidelity $1 - F\left(U_{\text{ideal}}, U_{\text{Magnus}}\left(\tau\right)\right)$, given by Eq.~\eqref{eq:F_analytical}, vanishes when $N_{\mathrm{d}}$ is odd because the coefficients in Eqs.~\eqref{eq:lambda_x_0}--\eqref{eq:lambda_z_0} satisfy $\lambda_{\mathrm{x}} = -\phi/2$, $\lambda_{\mathrm{y}}=0$, and $\lambda_{\mathrm{z}}=0$. 
		However, the numerical value $1 - F\left(U_{\text{ideal}}, U\right)$ remains nonzero, as shown in Fig.~\ref{fig:result:N}. 
		This discrepancy 
originates from the RWA employed to derive $1 - F\left(U_{\text{ideal}}, U_{\text{Magnus}}\left(\tau\right)\right)$.
		The fast-oscillating terms neglected under the RWA cause residual crosstalk in the numerical evolution, resulting in a nonzero value of the infidelity $1 - F\left(U_{\text{ideal}}, U\right)$.

	\subsection{Pulse width requirements for AIS} \label{sec:result:tau}
		Here, we present the pulse width requirements for AIS. 
		Figure~\ref{fig:result:tau}(a) shows the infidelity $1 - F_{\text{mean}}\left(U_{\text{ideal}}, U\right)$ as a function of the number of drive microwave tones $N_\mathrm{d}$ and normalized pulse width $\tau/\tau_0$. 
		Figure~\ref{fig:result:tau}(b) presents an enlarged view of Fig.~\ref{fig:result:tau}(a) in the range $0.9 \le \tau/\tau_0 \le 1.1$, corresponding to the region outlined by the orange dashed lines. 
Note that we set the pulse amplitude $\alpha$ according to the pulse width $\tau$ to satisfy the condition $\tau\alpha = \phi$, where $\phi$ is the desired rotation angle.
        Thus, if the interference from off-resonant microwave tones is negligible, 
        the fidelity should remain high regardless of the pulse width. 
		For $\tau = m\tau_0/2$, i.e., in the orthogonal and quasi-orthogonal cases, the infidelity decreases as $N_{\mathrm{d}}$ increases; thus, AIS has a beneficial effect. 
This result is consistent with the analytical predictions presented in Sec.~\ref{sec:method}.
		However, for $\tau \ne m\tau_0/2$, the infidelity remains almost constant over the entire range of $N_{\mathrm{d}}$. 
This is because, in this case, 
the unwanted terms in the coefficients $\lambda_{\mathrm{x}}$, $\lambda_{\mathrm{y}}$, and $\lambda_{\mathrm{z}}$ do not vanish, thereby limiting the effectiveness of AIS (see Appendix~\ref{sec:appendix:lambda}).

		This result significantly restricts the choices of $\tau$. 
		However, $\tau$ is already selected from the set $\left\{m\tau_0/2\right\}$ to achieve high fidelity in the underlying OFDM scheme~\cite{OFDM_gate}.  
		Therefore, this restriction does not constitute a significant limitation of AIS. 

\section{Compensating for the effects of fast-oscillating terms} \label{sec:result2:shift}
	In this section, we numerically demonstrate that fast-oscillating terms degrade the overall gate fidelity. 
	Although Eqs.~\eqref{eq:F_analytical}--\eqref{eq:lambda_z_0} show that the infidelity for $\tau = m\tau_0/2$ with 
odd $N_{\mathrm{d}}$ and $S_{\mathrm{d}} = 0$ is symmetric about $k_{\mathrm{q}} = 0$ under the RWA, this symmetry is broken beyond the RWA.
    In particular, the terms involving $2\omega_{\mathrm{q}, k_{\mathrm{q}}} + \Delta_{k_{\mathrm{d}} k_{\mathrm{q}}}$ in Eq.~\eqref{eq:HI} introduce slight asymmetry, leading to a degradation in the overall gate fidelity. 
	Consequently, we show that fine-tuning the drive frequency allocation can compensate for this effect and improve the overall gate performance.

	\subsection{Dependence on qubit frequency} \label{sec:result2:shift:omega_q}
		\begin{figure*}[tb]
			\begin{tabular}{cc}
				\begin{minipage}[t]{3.5in}
					\centering
					\includegraphics[scale=1]{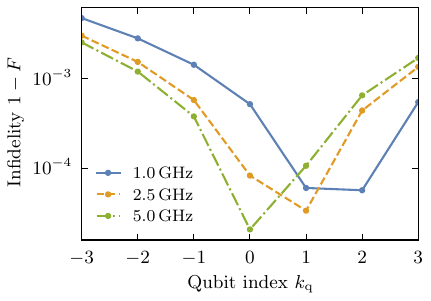}
				\end{minipage}
				\begin{minipage}[t]{3.5in}
					\centering
					\includegraphics[scale=1]{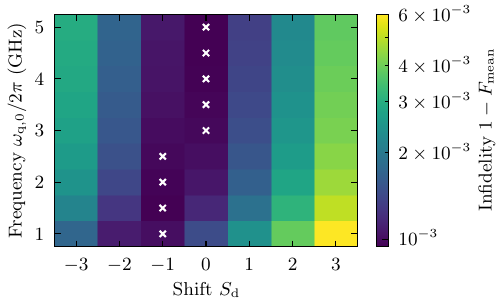}
				\end{minipage}
			\end{tabular}
			\begin{picture}(700, 0)
				\put(26, 140){(a)}
				\put(250, 140){(b)}
			\end{picture}
			\caption{(a) Infidelity $1 - F\left(U_{\text{ideal}}, U\right)$ as a function of the qubit index $k_{\mathrm{q}}$ for $S_{\mathrm{d}} = 0$,  
				where three curves are shown for $\omega_{\mathrm{q}, 0} / 2\pi = 1,\, 2.5$, and $5 \,\mathrm{GHz}$.  
				(b) Infidelity $1 - F_{\text{mean}}\left(U_{\text{ideal}}, U\right)$ as a function of the shift parameter $S_{\mathrm{d}}$ and the central qubit frequency $\omega_{\mathrm{q}, 0}$. 
				$F_{\text{mean}}\left(U_{\text{ideal}}, U\right)$ denotes the mean of $F\left(U_{\text{ideal}}, U\right)$ over all qubits $k_{\mathrm{q}} \in K_{\mathrm{q}} = \left\{-3, -2, -1, 0, 1, 2, 3\right\}$. 
				The parameters are $N_{\mathrm{q}} = 7$, $N_{\mathrm{d}} = 21$, $\Delta/2\pi = 10 \,\mathrm{MHz}$, $\phi = \pi / 2$, $\tau = \tau_0$, and $\alpha = -\phi/\tau$. 
				In (b), the local minima of infidelity for each $\omega_{\mathrm{q}, 0}$ are plotted with crosses.
				The operator $U$ is obtained numerically from the Hamiltonian in Eq.~\eqref{eq:H}. 
				}
			\label{fig:result:shift:omega_q}
		\end{figure*}
		
		In Fig.~\ref{fig:result:shift:omega_q}(a), we plot the infidelity $1 - F\left(U_{\text{ideal}}, U\right)$ as a function of the qubit index $k_{\mathrm{q}}$ for $\omega_{\mathrm{q}, 0} / 2\pi = 1$, $2.5$, and $5 \,\mathrm{GHz}$. 
        Here, as described in Sec.~\ref{sec:model:hamiltonian}, 
        we choose the drive frequencies such that $\omega_{\mathrm{d}, k_{\mathrm{d}}} = \omega_{\mathrm{q}, k_{\mathrm{q}}}$ for $k_{\mathrm{d}} = k_{\mathrm{q}}$. 
		We set $N_{\mathrm{q}} = 7$ and $N_{\mathrm{d}} = 21$ and place the microwave tones symmetrically with respect to the central qubit $k_{\mathrm{q}} = 0$, i.e., $S_{\mathrm{d}} = 0$. 
		The evolution operator $U$ is obtained numerically from the Hamiltonian in Eq.~\eqref{eq:H}. 
		Notably, the infidelity is asymmetric about $k_{\mathrm{q}} = 0$; the infidelity for negative $k_{\mathrm{q}}$ is larger than that for positive $k_{\mathrm{q}}$.
		Therefore, we expect the mean infidelity over $k_{\mathrm{q}} \in K_{\mathrm{q}}$ to decrease when the centroid of the microwave tones is shifted slightly toward lower frequencies. 
		In addition, this asymmetry increases as $\omega_{\mathrm{q}, 0}$ decreases; therefore, a larger shift is required for a smaller $\omega_{\mathrm{q}, 0}$. 
		
		Figure~\ref{fig:result:shift:omega_q}(b) shows the infidelity $1 - F_{\text{mean}}\left(U_{\text{ideal}}, U\right)$
		as a function of the shift parameter $S_{\mathrm{d}}$ and $\omega_{\mathrm{q}, 0}/2\pi$ for $N_{\mathrm{q}}=7$ and $N_{\mathrm{d}}=21$. 
        The microwave tones are symmetrically allocated with respect to $k_{\mathrm{d}} = S_{\mathrm{d}}$ [see Fig.~\ref{fig:K}(c) for an example of $K_{\mathrm{d}}$ with $N_{\mathrm{d}} = 9$ and $S_{\mathrm{d}} = -1$]. 
The crosses correspond to the minimum infidelity for each value of $\omega_{\mathrm{q},0}$. 
		The shifts yielding the minimum infidelity move to the left as $\omega_{\mathrm{q}, 0}$ decreases. 
		Specifically, for $\omega_{\mathrm{q}, 0} / 2\pi = 1$, $1.5$, and $2\,\mathrm{GHz}$, 
        we observe the minimum infidelity at $S_{\mathrm{d}} = -1$, whereas it is at $S_{\mathrm{d}} = 0$ for $\omega_{\mathrm{q}, 0} / 2\pi \geq 2.5$ GHz.
		For $\omega_{\mathrm{q}, 0} / 2\pi = 1\,\mathrm{GHz}$, we observe a reduction in the mean infidelity $1 - F_{\text{mean}}\left(U_{\text{ideal}}, U\right)$ from $1.45 \times 10^{-3}$ at $S_{\mathrm{d}} = 0$ to $1.01 \times 10^{-3}$ at $S_{\mathrm{d}} = -1$, a $30.3 \%$ decrease. 

	\subsection{Dependence on the number of drive tones} \label{sec:result2:shift:Nd}
		\begin{figure*}[tb]
			\begin{tabular}{cc}
				\begin{minipage}[t]{3.5in}
					\centering
					\includegraphics[scale=1]{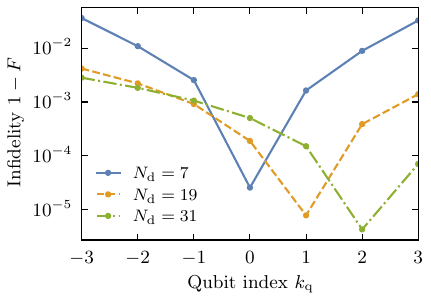}
				\end{minipage}
				\begin{minipage}[t]{3.5in}
					\centering
					\includegraphics[scale=1]{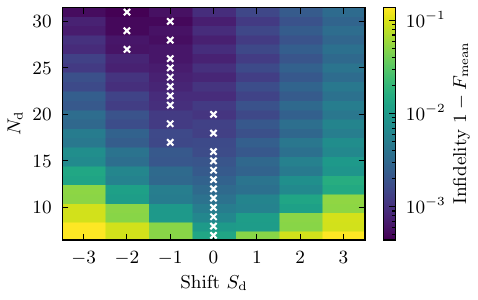}
				\end{minipage}
			\end{tabular}
			\begin{picture}(700, 0)
				\put(26, 140){(a)}
				\put(270, 140){(b)}
			\end{picture}
			\caption{(a) Infidelity $1 - F\left(U_{\text{ideal}}, U\right)$ as a function of the qubit index $k_{\mathrm{q}}$ for $S_{\mathrm{d}} = 0$,  
				where three curves are shown for $N_{\mathrm{d}} = 7,\, 19$, and $31$.  
				(b) Infidelity $1 - F_{\text{mean}}\left(U_{\text{ideal}}, U\right)$ as a function of the shift parameter $S_{\mathrm{d}}$ and the number of drive microwave tones $N_{\mathrm{d}}$. 
				$F_{\text{mean}}\left(U_{\text{ideal}}, U\right)$ is the mean of $F\left(U_{\text{ideal}}, U\right)$ over all qubits $k_{\mathrm{q}} \in K_{\mathrm{q}} = \left\{-3, -2, -1, 0, 1, 2, 3\right\}$. 
				The parameters are $N_{\mathrm{q}} = 7$, $\omega_{\mathrm{q}, 0} / 2\pi = 1.5 \,\mathrm{GHz}$, $\Delta/2\pi = 10 \,\mathrm{MHz}$, $\phi = \pi / 2$, $\tau = \tau_0$, and $\alpha = -\phi/\tau$. 
				In (b), local minima of infidelity for each $N_{\mathrm{d}}$ are plotted with crosses. 
				The evolution operator $U$ is obtained numerically from the Hamiltonian in Eq.~\eqref{eq:H}. 
				}
			\label{fig:result:shift:N}
		\end{figure*}
		
		The number of drive microwave tones $N_{\mathrm{d}}$ also affects the optimal value of the shift $S_{\mathrm{d}}$.  
		Figure~\ref{fig:result:shift:N}(a) shows the infidelity as a function of the qubit index $k_{\mathrm{q}}$ for $N_{\mathrm{d}} = 7$, $19$, and $31$. 
We set $N_{\mathrm{q}} = 7$ and place the microwave tones symmetrically with respect to the central qubit $k_{\mathrm{q}} = 0$, as in Fig.~\ref{fig:result:shift:omega_q}(a).
		The asymmetry of the infidelity about $k_{\mathrm{q}} = 0$ increases for larger $N_{\mathrm{d}}$. 
		This behavior arises because the contributions of the terms neglected under the RWA (associated with $2\omega_{\mathrm{q}, k_{\mathrm{q}}} + \Delta_{k_{\mathrm{d}} k_{\mathrm{q}}}$) accumulate over all the drive microwave tones. 
		Therefore, a larger shift is required for a larger $N_{\mathrm{d}}$ to compensate for the increased asymmetry, thereby improving the mean fidelity over all qubits. 
		
		Figure~\ref{fig:result:shift:N}(b) shows the mean gate infidelity $1 - F_{\text{mean}}\left(U_{\text{ideal}}, U\right)$ as a function of the shift $S_{\mathrm{d}}$ and the number of drive microwave tones $N_{\mathrm{d}}$. 
		The optimal shift for each $N_{\mathrm{d}}$, plotted with a cross, tends to move to the left as $N_{\mathrm{d}}$ increases. 
Notably, for $N_{\mathrm{d}} = 31$, we obtain a mean infidelity of $4.39 \times 10^{-4}$ at $S_{\mathrm{d}} = -2$, a $52.3 \%$ reduction compared with $9.22 \times 10^{-4}$ at $S_{\mathrm{d}} = 0$. 

\section{Discussion}
	In Sec.~\ref{sec:result2:shift}, we have shown that the infidelity can be reduced by considering the effects neglected under the RWA. 
	When $N_{\mathrm{d}} = 1$, which is a typical value for conventional one-to-one wiring systems, 
	the contributions of the fast-oscillating terms to the fidelity are negligible because they are several orders of magnitude smaller than those of the resonant terms. 
	However, as $N_{\mathrm{d}}$ increases, these contributions accumulate over all the drive microwave tones. Therefore, their impact on gate performance must be considered when adopting AIS, which uses a large number of drive microwave tones. 

    Although we assumed rectangular envelopes in this paper,
    the AIS framework is compatible with non-square envelopes in the sense that fidelity can be improved by adding off-resonant tones symmetrically relative to the central qubit's frequency at regular intervals.
    This is because AIS compensates for the effects of off-resonant tones by balancing the drive frequency allocation. 
    Nevertheless, several properties, including the scaling of the infidelity and the pulse width requirements, may differ depending on the pulse envelope. 
	
	We have not considered leakage outside the computational subspace to isolate and clarify the essential physics of the AIS method. 
    However, additional off-resonant drive tones can increase leakage because their spectra can overlap with leakage transition frequencies. 
	One possible measure to suppress such leakage is to replace the edges 
	of the rectangular pulse envelope with a raised cosine envelope and employ derivative removal by adiabatic gate (DRAG) pulses~\cite{DRAG1, DRAG2}. 
Furthermore, the feasible values of $N_{\mathrm{d}}$ and $N_{\mathrm{q}}$ are limited by the frequency spacing $\Delta$ to avoid spectral overlap with the leakage transition frequency.
	This constraint represents a limitation of the proposed approach.

    In practical implementations, increasing $N_{\mathrm{d}}$ increases the total microwave power delivered through the shared control line. 
    Consequently, $N_{\mathrm{d}}$ is also limited by hardware constraints, such as the dynamic range of the control electronics and the cooling capacity of the cryostat. 
    In addition, the present AIS implementation relies on precise frequency and phase relationships among microwave tones. 
    Therefore, AIS may be sensitive to parameter variations, such as qubit frequency drift and microwave-tone phase offsets. 
    This sensitivity may require stringent specifications for the control electronics and qubits. 
    Developing AIS frameworks that are robust against such parameter variations is an interesting direction for future study.

\section{Conclusion}
	In this study, we developed an active interference suppression method for simultaneous gate operations based on the OFDM scheme. 
	We have demonstrated a reduction in the infidelity of single-qubit gates by deliberately incorporating off-resonant orthogonal or quasi-orthogonal microwave tones. 
Notably, we have analytically and numerically shown that the infidelity scales as $N_{\mathrm{d}}^{-2}$, the inverse square of the number of incorporated microwave tones.

Further, we have shown that fast-oscillating terms neglected under the RWA affect gate fidelity. 
	By shifting the centroid of the off-resonant microwave tones in the frequency domain, we mitigated this effect and achieved higher fidelity.

    In Ref.~\cite{OFDM_gate}, it has already been shown that the incorporation of appropriately selected off-resonant tones can suppress interference. 
However, Ref.~\cite{OFDM_gate} primarily focused on the utilization of spectral zero crossings under the assumption of $N_{\mathrm{d}} = N_{\mathrm{q}}$. 
    Thus, it lacked a quantitative investigation of interference suppression using off-resonant tones.
    In contrast, the present work extends the analysis to the regime of $N_{\mathrm{d}} > N_{\mathrm{q}}$ and investigates detailed characteristics that go beyond the scope of the previous study. 
    
    We have focused on simultaneous gate operations based on the FDM scheme. 
    Nevertheless, we expect that the AIS concept can be applied to related control problems, such as selective excitation~\cite{selective_excitation} and microwave crosstalk mitigation~\cite{microwave_crosstalk_APL}.
	
\begin{acknowledgments}
	This study used QuTiP~\cite{qutip5} for numerical calculations.	
\end{acknowledgments}

\appendix

\section{Origin of the reduced effectiveness of AIS for $\tau \ne m\tau_0/2$} \label{sec:appendix:lambda}
	As shown in Sec.~\ref{sec:result:tau}, the effectiveness of AIS decreases for $\tau \ne m\tau_0/2$.
	In this appendix, we examine this behavior through the coefficients $\lambda_{\mathrm{x}}$, $\lambda_{\mathrm{y}}$, and $\lambda_{\mathrm{z}}$ introduced in Eqs.~\eqref{eq:Omega_1_1} and \eqref{eq:Omega_2_1}.

	The first and second terms of the Magnus series, $\Omega_1\left(t\right)$ and $\Omega_2\left(t\right)$, introduced in Eq.~\eqref{eq:Magnus}, are given by 
	\begin{align}
		\label{eq:Omega_1}
		\Omega_1\left(t\right) &\equiv \int_{0}^{t} H_{\mathrm{I}, k_{\mathrm{q}}}\left(t_1\right) \mathrm{d}t_1 , 
	\end{align}
	\begin{align}
		\label{eq:Omega_2}
		\Omega_2\left(t\right) &\equiv -\frac{i}{2} \int_{0}^{t} \int_{0}^{t_1} \left[ H_{\mathrm{I}, k_{\mathrm{q}}}\left(t_1\right), H_{\mathrm{I}, k_{\mathrm{q}}}\left(t_2\right) \right] \mathrm{d}t_2 \mathrm{d}t_1.
	\end{align}
	Substituting Eq.~\eqref{eq:HI_RWA} into Eqs.~\eqref{eq:Omega_1} and~\eqref{eq:Omega_2}, we obtain $\lambda_{\mathrm{x}}$, $\lambda_{\mathrm{y}}$, and $\lambda_{\mathrm{z}}$ for arbitrary $\tau$~\cite{OFDM_gate}: 
	\begin{align}
		\label{eq:lambda_x_general}
		\lambda_{\mathrm{x}} &=  
		\frac{\alpha\tau}{2}  \left[ 1 +  \sum_{\substack{k_{\mathrm{d}} \in K_{\mathrm{d}} \\ k_{\mathrm{d}} \ne k_{\mathrm{q}}}} \mathrm{sinc} (\Delta_{k_{\mathrm{d}}k_{\mathrm{q}}} \tau) \right],
	\end{align}
	\begin{align}
		\label{eq:lambda_y_general}
		\lambda_{\mathrm{y}} &=  -
		\frac{\alpha \tau}{2} 
		\sum_{k_{\mathrm{d}} \in \gamma} \sin \left(\frac{\Delta_{k_{\mathrm{d}} k_{\mathrm{q}}} \tau}{2}\right) 
		\mathrm{sinc} \left(\frac{\Delta_{k_{\mathrm{d}} k_{\mathrm{q}}} \tau}{2}\right), 
	\end{align}
	\begin{widetext}
		\begin{align}
			\label{eq:lambda_z_general}
			\lambda_{\mathrm{z}} &=
			 \frac{\alpha^2 \tau}{4} 
			\left[ \sum_{k_{\mathrm{d}} \in \gamma} \frac{ \cos(\Delta_{k_{\mathrm{d}} k_{\mathrm{q}}} \tau ) -  \frac{3}{2}\mathrm{sinc}(\Delta_{k_{\mathrm{d}} k_{\mathrm{q}}} \tau)}{\Delta_{k_{\mathrm{d}} k_{\mathrm{q}}}}
+ \sum_{k_{\mathrm{d}} \in \gamma}
			\sum_{\substack{j \in K_{\mathrm{d}} \\ j \ne k_{\mathrm{q}}}} \left\{ 
			\frac{\mathrm{sinc}\left(\Delta_{jk_{\mathrm{q}}}\tau \right)}{\Delta_{k_{\mathrm{d}} k_{\mathrm{q}}}} + \frac{\mathrm{sinc}\left[\left(\Delta_{k_{\mathrm{d}} k_{\mathrm{q}}} + \Delta_{j k_{\mathrm{q}}}\right)\tau\right]}{2\Delta_{j k_{\mathrm{q}}}}
			\right\} \right. \nonumber \\
			&\qquad\qquad\qquad\qquad\left.- \sum_{k_{\mathrm{d}} \in K_{\mathrm{d}}}
			\sum_{j \in \gamma} \frac{\mathrm{sinc}\left[\left(\Delta_{k_{\mathrm{d}}k_{\mathrm{q}}} + \Delta_{jk_{\mathrm{q}}}\right)\tau\right]}{2\Delta_{jk_{\mathrm{q}}}}
- \sum_{k_{\mathrm{d}} \in \gamma}
			\sum_{\substack{j \in K_{\mathrm{d}} \\ j \ne k_{\mathrm{q}} \\ j \ne k_{\mathrm{d}}}} \left\{ \frac{1}{\Delta_{k_{\mathrm{d}}k_{\mathrm{q}}}} + \frac{1}{\Delta_{jk_{\mathrm{q}}}} \right\}\frac{\mathrm{sinc}(\Delta_{k_{\mathrm{d}}j}\tau)}{2}\right]. 
		\end{align}
	\end{widetext}
For $\tau \ne m\tau_0/2$, the terms proportional to the sinc function remain nonzero. 
	In particular, the remaining $\mathrm{sinc} (\Delta_{k_{\mathrm{d}}k_{\mathrm{q}}} \tau)$ terms appearing in the summation in Eq.~\eqref{eq:lambda_x_general} cannot be suppressed using the AIS scheme, thereby reducing its effectiveness. 
	By contrast, these terms vanish for $\tau = m\tau_0/2$ because $\sin\left(n \Delta \tau_0\right) = 0$ and $\sin\left(n\frac{\Delta\tau_0}{2}\right) = 0$ for all integers $n$. 
	The explicit forms of $\lambda_{\mathrm{x}}$, $\lambda_{\mathrm{y}}$, and $\lambda_{\mathrm{z}}$ for $\tau = m\tau_0/2$, shown in Eqs.~\eqref{eq:lambda_x_0}--\eqref{eq:lambda_z_0}, are obtained by substituting $\alpha = -\phi/\tau$ and $\tau = m\tau_0/2$ into Eqs.~\eqref{eq:lambda_x_general}--\eqref{eq:lambda_z_general}.

\bibliography{bib/quantum_algorithm, bib/PQC, bib/reduce_control_line, bib/control, bib/others, bib/quantum_device, bib/FTQC}

\end{document}